\documentclass{appolb}
\usepackage{graphicx}

\begin{document}
\title{Latest developments in anisotropic hydrodynamics%
\thanks{Presented at {\it Excited QCD 2015}, 8-14 March 2015
Tatranska Lomnica, Slovakia.}%
}
\author{LeonardoTinti
\address{Institute of Physics, Jan Kochanowski University, PL-25406~Kielce, Poland}
}
\maketitle

\begin{abstract}

We discuss the leading order of anisotropic hydrodynamics expansion. It has already been shown that in the (0+1) and (1+1)-dimensional cases it is consistent with the second order viscous hydrodynamics, and it provides a striking agreement with the exact solutions of the Boltzmann equation. Quite recently, a new set of equations has been proposed for the leading order of anisotropic hydrodynamics, which is consistent with the second order viscous hydrodynamics in the most general (3+1)-dimensional case, and does not require a next-to-leading treatment for describing pressure anisotropies in the transverse plane.

\end{abstract}


\section{Introduction}

Relativistic hydrodynamics plays an essential role in relativistic heavy-ion collisions, see for instance Refs.~\cite{
Muronga:2003ta,
Luzum:2008cw,
Schenke:2011tv,
Shen:2011,
Bozek:2012qs,
Denicol:2012cn}. Early calculations were based on ideal hydrodynamics, nowadays, however, viscous hydrodynamics is preferred. Both because it provides a better description of the data and because of general arguments that the fluid shear viscosity cannot be zero, the latter fact following from quantum mechanical considerations~\cite{micro} as well as from the AdS/CFT correspondence~\cite{ADS/CFT}. Despite its obvious success, there are still fundamental issues with the ordinary expansion. In the second order viscous hydrodynamics, momentum anisotropies and pressure corrections are treated like small perturbations, however in heavy-ion collision conditions they are often very large. 

A new approach to address these problems is {\it anisotropic hydrodynamics} (aHydro)~\cite{
Martinez:2012tu,
Ryblewski:2012rr,
Ryblewski:2013jsa,
Florkowski:2012ax,
Florkowski:2014txa,
Florkowski:2014sfa}, where the large momentum anisotropies, and hence large pressure corrections, are treated in a non perturbative way.  Anisotropic hydrodynamics is a reorganization of the relativistic hydrodynamics expansion, done around a non-isotropic background. The leading order already contains substantial momentum-space anisotropies, which can reproduce the large pressure anisotropies generated in ultrarelativistic heavy-ion collisions. 

\section{The hydrodynamics expansion}

\begin{figure*}[t]
\begin{center}
\includegraphics[angle=0,width= \textwidth]{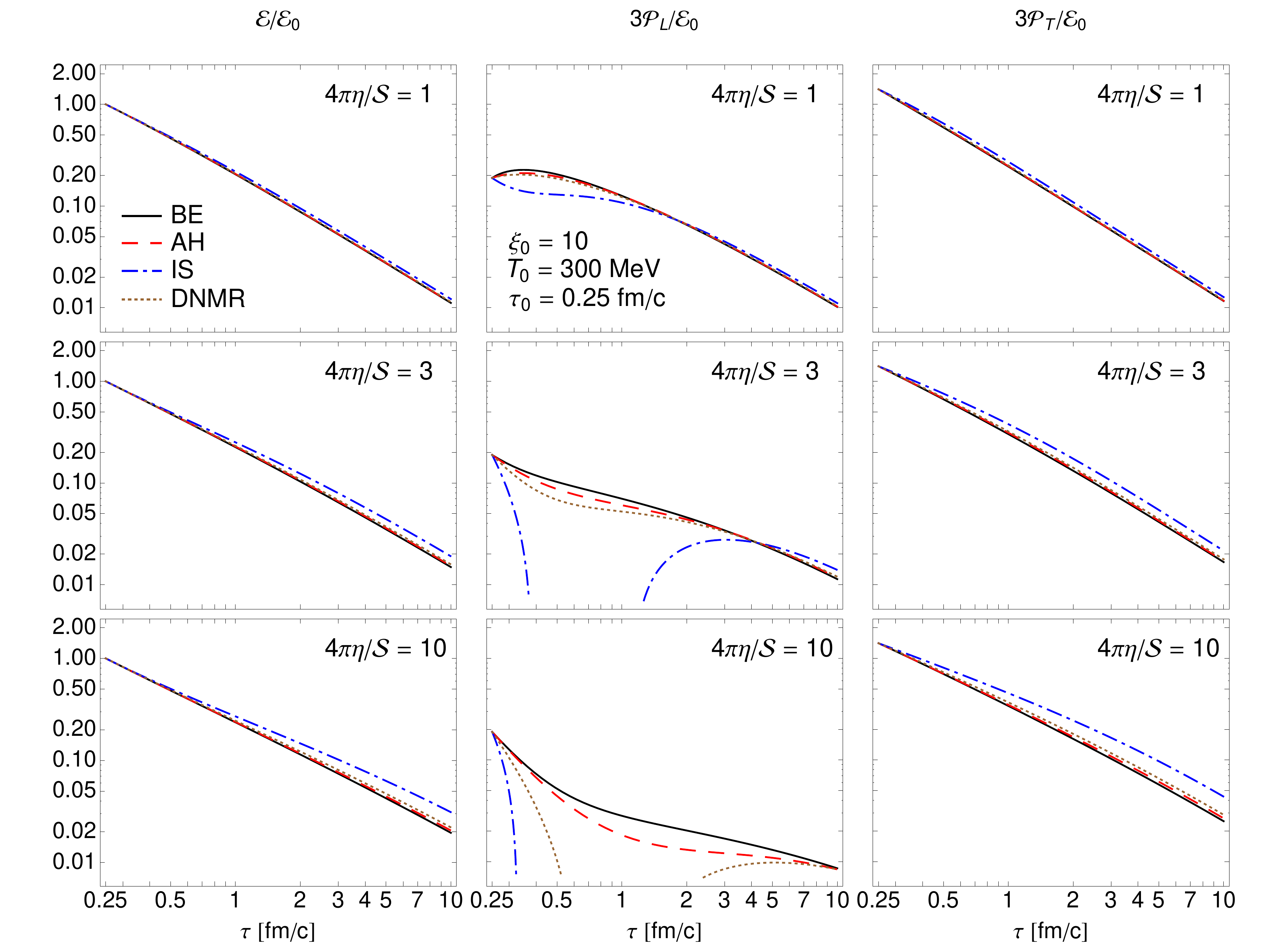}
\end{center}
\caption{(Color online) Comparison of anisotropic hydrodynamics and second-order viscous hydrodynamics with the exact solution of the Boltzmann equation, for energy density, longitudinal pressure, and transverse pressure evolution(figure taken from \cite{Florkowski:2013lya}).
}
\label{fig:PLPTE_300_10}
\end{figure*}

The most common assumption for deriving hydrodynamics from relativistic kinetic theory is that the particle distribution function $f(x,p)$ is very close to local equilibrium. Ignoring conserved charges and assuming the Boltzmann phase-space distribution for equilibrium we may write
\begin{equation}
 f(x,p) = f_{\rm eq.}(x,p) + \delta f(x,p), \qquad f_{\rm eq.}(x,p) = k \exp\left[-\frac{p\cdot U(x)}{T(x)} \right],
\label{hydro_expansion}
\end{equation}
with $T$ and $U^\mu$ being the effective temperature and the fluid four velocity, respectively. The leading order in the expansion~(\ref{hydro_expansion}), $f_{\rm eq.}$, describes the perfect fluid, while the viscous correction depends only on $\delta f$, which is treated as a small perturbation. However, when we consider an (almost) boost invariant flow like the one we expect in the early stages of heavy-ion collisions, the four velocity gradients are inversely proportional to the proper time. Therefore, the pressure corrections become close to the equilibrium pressure, questioning the validity of the perturbative treatment.

The main feature of anisotropic hydrodynamics is to treat the large momentum anisotropy in a non perturbative way starting from the leading order, namely, we write

\begin{equation}
 f(x,p) = f_{\rm aniso.}(x,p) +\delta \tilde{f}(x,p).
\label{aniso_exp}
\end{equation}
In this way, the deviation $\delta\tilde{f}$ from the (non isotropic and dissipative) background $f_{\rm aniso.}$ can be small enough to justify a perturbative treatment. The first formulation of aHydro used  the point dependent version of the Romatschke-Strickland form (presented in~\cite{Romatschke:2003ms}) for the leading order of the anisotropic expansion, which in the local rest frame (LRF) reads

\begin{equation}
 f_{\rm aniso.}(x,p) = k\exp\left[ -\frac{1}{\Lambda(x)}\sqrt{ \frac{}{} p^2_T +\zeta(x) p_L^2} \right].
\label{RS}
\end{equation}
Here $\Lambda$  is the momentum scale\footnote{The effective temperature $T$ is defined using the Landau matching, and it is different from $\Lambda$ in general.}, $p_T$ and $p_L$ are the transverse and longitudinal momenta, and $\zeta$ is the anisotropy parameter. In order to locally conserve energy and momentum, one has to use the first moment of the Boltzmann equation, which provides four independent equations.  In order to close the system of equations for the leading order of anisotropic hydrodynamics, at first the particle creation equation has been used, namely, the zeroth moment of the Boltzmann equation.

For a longitudinally boost invariant and transversely homogeneous system there is an exact solution of the relativistic Boltzmann equation~\cite{Florkowski:2013lya}. We show in Fig.~\ref{fig:PLPTE_300_10} one of the plots from~\cite{Florkowski:2013lya}. The comparison is done between the exact solution (BE), Israel-Stewart theory (IS), the new formulation of the second-order viscous hydrodynamics presented in~\cite{Denicol:2012cn} (DNMR), and anisotropic hydrodynamics (AH). Anisotropic hydrodynamics is always very close to the exact solution, while IS is providing unphysical vanishing longitudinal pressure ${\cal P}_L$, and significant deviations from the exact evolution of the transverse pressure ${\cal P}_T$. In the most extreme case (large shear viscosity over entropy ratio $\eta/{\cal S}$), even the DNMR approach is not reliable.

\section{Improvement the leading order of anisotropic expansion}

\begin{figure}[ht]
\begin{minipage}[b]{0.47 \linewidth}
\centering
\includegraphics[angle=0,width=1.1 \textwidth]{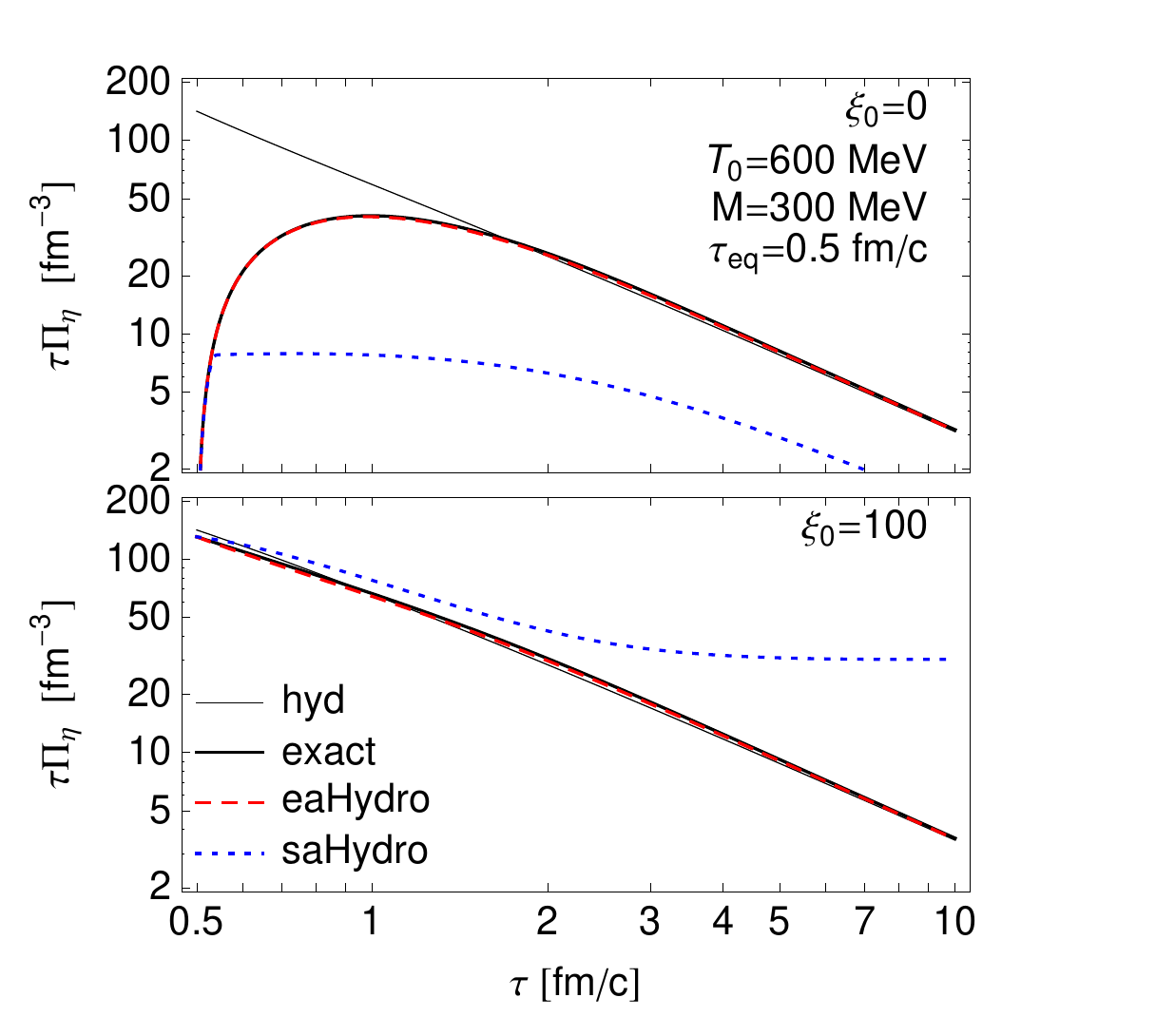}
\end{minipage}
\hspace{1.0cm}
\begin{minipage}[b]{0.47\linewidth}
\centering
\includegraphics[angle=0,width=1.1 \textwidth]{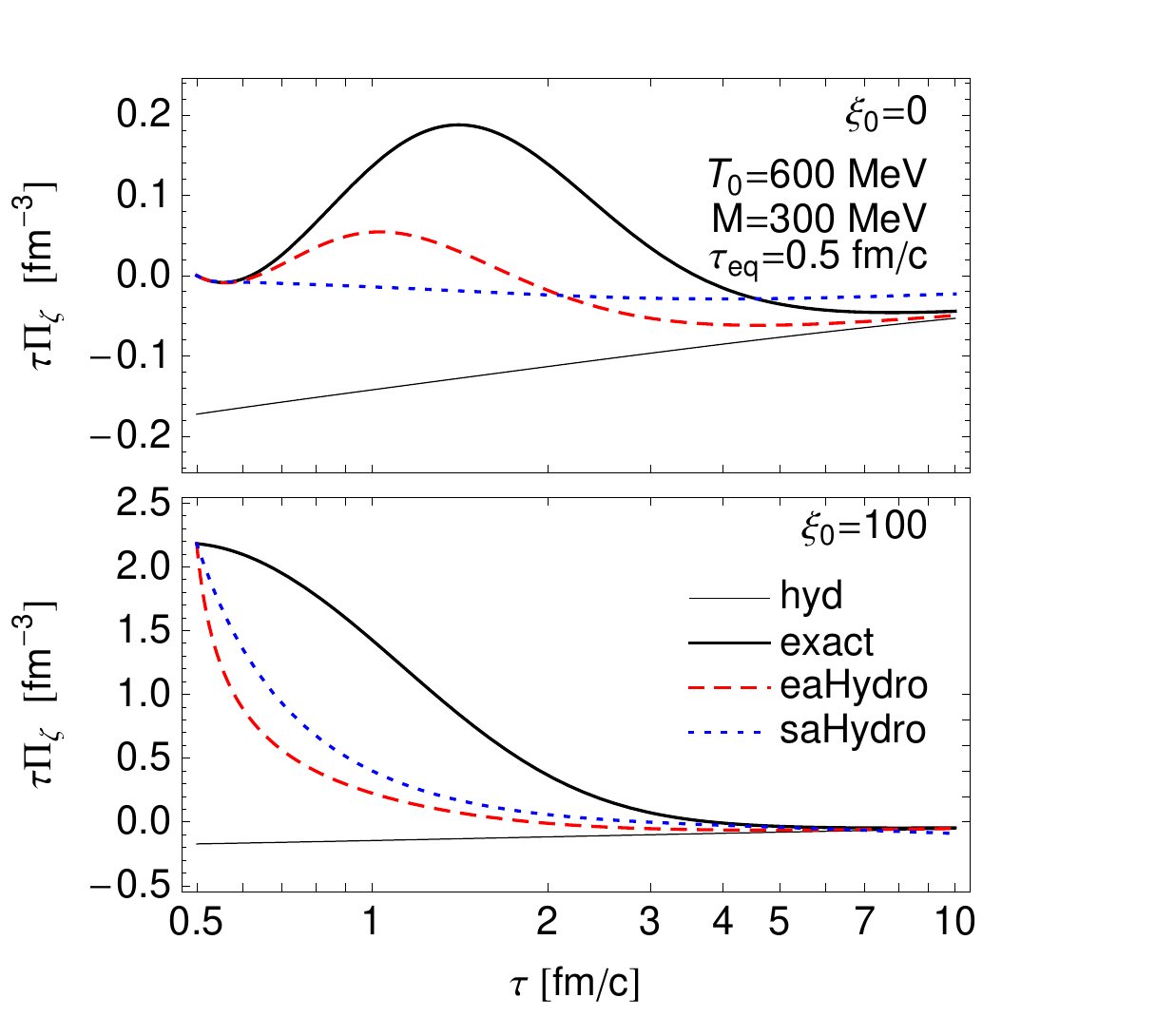}
\end{minipage}
\caption{(Color online) Time dependence of shear and bulk viscous pressure multiplied by $\tau$ (figure taken from ~\cite{Florkowski:2014nc}).}
\label{figs:shear_bulk}
\end{figure}

The anisotropic background~(\ref{RS}) takes into account differences between the longitudinal pressure ${\cal P}_L$ and the transverse pressure ${\cal P}_T$, only. However if there is a non-vanishing radial flow we expect anisotropies even in the transverse plane. As the system evolves toward equilibrium, these corrections become as important as the longitudinal ones. One way to handle the non trivial transverse dynamics is to treat $\delta\tilde{f}$ in the anisotropic expansion~(\ref{aniso_exp}) in a perturbative way~\cite{Bazow:2013ifa, Bazow:2015cha}. Alternatively,  we propose here to include most of the dynamic effects connected with anisotropy in the leading order itself.

In Ref.~\cite{Tinti:2014conf} we extended the formalism of anisotropic hydrodynamics to the (1+1)-dimensional case. In order to obtain a closed set of equations, we used the second moment of the Boltzmann equation, in addition to the energy and momentum conservation. We proved that these equations reduce to the Israel-Stewart equations in the close to equilibrium limit, where we know that the second-order viscous hydrodynamics is justified. Later, we compared this new set of equations with the solution of the Boltzmann equation and the original prescription for anisotropic hydrodynamics~\cite{Florkowski:2014nc}. There is a large improvement in the agreement with the exact solution, especially for massive particles. In Fig.~\ref{figs:shear_bulk} we show the comparison between the new formulation (eaHydro) and the original one (saHydro). The shear evolution $\tau\Pi_\eta$ is very well reproduced, while the bulk evolution $\tau\Pi_\zeta$ still shows some deviations from the exact solution. Note that $\tau$ is the (longitudinal) proper time, ${\cal P}_{\rm eq.}$ is the equilibrium pressure, $ \Pi_\eta = \frac{2}{3}\left( {\cal P}_T -\frac{}{}{\cal P}_L\right)$, and $\Pi_\zeta = \frac{1}{3}\left(\frac{}{} 2{\cal P}_T + {\cal P}_L -3{\cal P}_{\rm eq.} \right)$.

In~Ref.~\cite{Nopoush:2014nc}  an explicit degree of freedom to take into account the bulk evolution has been proposed, allowing for a better reproduction of the isotropic pressure corrections. Very recently~\cite{Tinti:2014yya} the leading order of anisotropic hydrodynamics has been extended to the most general flow, the $(3+1)$-dimensional case. There are no restrictions from boost invariance or cylindrical symmetry. We used a generalized Romatschke-Strickland form for the anisotropic background

\begin{equation}
 f_{\rm aniso.} = k \exp\left[ -\frac{1}{\lambda}\sqrt{(p\cdot U)^2 +p_\mu\left( \frac{}{} \xi^{\mu\nu} -\phi \Delta^{\mu\nu} \right)p_\nu} \right].
\end{equation}
The anisotropy tensor $\xi^{\mu\nu}$ is space-like and traceless, the scalar $\phi$ is taking into account the bulk dynamics, and $\lambda$ and $U$ are still, respectively, the momentum scale and the four-velocity.

We closed the system of equations using the first moment of the Boltzmann equation, and we used a geometrical argument for choosing the remaining equations from the zeroth and second moments. Finally, we proved that this generalization is fully consistent with the second order hydrodynamics in the limit of small deviation from local equilibrium.

%
%

\section{Conclusions}

Anisotropic hydrodynamics is a reorganization of the hydrodynamic expansion around a non-isotropic background. The leading order already provides large longitudinal pressure corrections, justifying the perturbative treatment of the next to leading order in heavy ion collisions. The original prescription has a striking agreement with the exact solutions of the Boltzmann equation in $(0+1)$ dimensions, but does not take into account pressure anisotropies in the transverse plane, therefore, requiring a next to leading order treatment in presence of transverse expansion. The first extension of the original treatment, allowing for cylindrically symmetric radial expansion, improved the agreement with the exact solutions in the Bjorken flow limit, remaining consistent with viscous hydrodynamics for small deviations from local equilibrium.

The newest prescription for the leading order is still reproducing second order viscous hydrodynamics in the close to equilibrium limit, and it is valid for the full $(3+1)$-dimensional expansion, not requiring a next to leading order treatment for including all transverse pressure corrections.

\section*{Acknowledgements}

This work has been supported by Polish National Science Center grant No. DEC-2012/06/A/ST2/00390.

\end{document}